\documentclass[superscriptaddress,prl,amsmath,twocolumn,amssymb]{revtex4}

\usepackage{dcolumn}
\usepackage{amsfonts}
\usepackage{amssymb}
\usepackage{amsmath}
\usepackage{amsthm}
\usepackage{latexsym}
\usepackage{epsfig}
\usepackage{color}
\usepackage{lscape}
\usepackage{multirow}
\usepackage{braket}
\usepackage{bbm}
\usepackage{mathrsfs}
\usepackage{fancyhdr}
\usepackage{lastpage}
\usepackage{tikz}
\usetikzlibrary{arrows}
\usetikzlibrary{shapes,fit}
\usepackage[vcentermath]{youngtab}

\newcommand{\be}{\begin{equation}}
\newcommand{\ee}{\end{equation}}
\newcommand{\ba}{\begin{array}}
\newcommand{\ea}{\end{array}}
\newcommand{\bea}{\begin{eqnarray}}
\newcommand{\eea}{\end{eqnarray}}

\def\Id{1\!\mathrm{l}}

\newcommand{\mr}[1]{\mathrm{#1}}
\def\Id{1\!\mathrm{l}}

\begin{document}

\title{Postselective quantum interference of distinguishable particles}

\author{Peter S. Turner}
\affiliation{School of Physics and Department of Electrical and Electronic Engineering, University of Bristol,\\ HH Wills Laboratory, Tyndall Avenue, Bristol BS8 1TL, United Kingdom}
\email{peter.turner@bristol.ac.uk}

\date{\today}

\begin{abstract}
We show that it is possible for completely distinguishable particles to interfere postselectively without operating on, or indeed having any knowledge of, the distinguishing degree of freedom.
In particular, we find a family of three-mode spatial interferometers that, upon inputting, say, a red photon in port 1 and a green photon in port 2, produce a state such that when vacuum is detected at one output, the two photons in the other two outputs will pass a HOM test, despite their frequency degree of freedom remaining untouched.
In doing so we develop a general approach to distinguishability based on the Schmidt decomposition between particles' ``System'' and ``Label'' degrees of freedom, corresponding to what has been called \emph{unitary-unitary duality} in many-body physics.
This also gives insight into the relationship between the first and second quantized pictures, which is useful in bringing other quantum information concepts across from the former to the latter.
\end{abstract}

\maketitle

Indistinguishability is arguably the defining feature of quantum mechanics.
Classic experiments such as the double slit lead us to believe that when there does not exist a measurement that distinguishes between two physical processes the probabilities for their outcomes exhibit interference, and that this interference is suppressed once distinguishability is introduced~\cite{ClassicText}.
It is a workhorse for demonstrating ``quantumness'', with many applications across a variety of fields; for example, quantum information science makes use of such interference as a signature for validating cryptographic keys~\cite{Crypto}, to sample probability distributions more efficiently than thought classically possible~\cite{AA}, and for universal quantum computation~\cite{KLM}.

Perhaps the best known example is the Hong-Ou-Mandel (HOM) effect~\cite{HOM}, where two photons are input into the two arms of a balanced beamsplitter.
When the photons are distinguishable, in this case by their arrival times, coincidence counts at the output are in agreement with classical predictions, interpreted as the sum of two distinguishable processes: both photons either transmit or both reflect.
When the timing degree of freedom is manipulated so as to make their arrival times the same, the photons become indistinguishable and quantum interference occurs, ideally yielding a coincidence rate of zero -- the celebrated HOM ``dip''.
This effect has been confirmed countless times, with near 100\% visibility shown in integrated photonics experiments~\cite{BestDip}.

We will use the HOM scenario extensively to motivate our model of distinguishability, which will lead us to consider quantum interference of distinguishable particles.
Each HOM photon has two pertinent degrees of freedom: one is spatial, namely the interferometer arms, and the other is temporal, namely the arrive times.
For practical applications, here we will be interested in the case where it is only the spatial degree of freedom over which we have control (via interferometry), and so we consider this the ``System'' degree of freedom.
We interpret the temporal degree of freedom as a ``Label''; when two particles share the same quantum state of their Label degree of freedom they are said to be indistinguishable, whereas when their Label states are orthogonal they are distinguishable.
There is a continuous range of partial distinguishability~\cite{Shchesnovich}, quantified by the overlaps between Label states.
The model makes no assumptions about the physical nature of either of these degrees of freedom -- it does not matter if the Label space is finite or infinite dimensional, overlaps can be computed in a Hilbert space of any dimensionality in the usual way.
In order to discuss distinguishability all we need is for the Label space to have as many dimensions as particles, so that they can all be labelled orthogonally.
For example, for $N$ spectrally distinguishable photons we need only consider the $N$ dimensional subspace of all square integrable functions spanned by the photons' spectral envelopes.

We start by modelling the HOM scenario, with two spatial System modes $s=1,2$ and two photons, requiring two orthogonal Label states for which we'll use two frequencies $l=R,G$, though this could just as easily be timing, polarization, etcetera.
Photon creators $\hat{a}^\dag_{s l}$\footnote{For example $\hat{a}^\dag_{s l} = \int \mathrm{d}\omega \, \hat{a}^\dag_{s}(\omega) f_l(\omega)$ where $f_l$ is a spectral envelope function indexed by $l$.} give rise to Fock states that we can write as arrays where rows correspond to System modes and columns to Label modes.
Canonical examples of indistinguishable\footnote{It is tempting to think that the two particles can be distinguished at this point by their System state, i.e. spatial mode 1 or 2; however, such a measurement would establish only that there was a particle in each mode, not which particle was in which mode -- such a distinction requires a Label degree of freedom.} and distinguishable two photon states are
\begin{align}
\ket{\psi_\mr{i}} &= \hat{a}^\dag_{1 R} \hat{a}^\dag_{2 R} \ket{\mr{vac}} = \Ket{\begin{matrix} 1&0\\1&0 \end{matrix}} , \label{eq:i}\\
\ket{\psi_\mr{d}} &= \hat{a}^\dag_{1 R} \hat{a}^\dag_{2 G} \ket{\mr{vac}} = \Ket{\begin{matrix} 1&0\\0&1 \end{matrix}} . \label{eq:d}
\end{align}
An interferometer acts only on the System, corresponding to a unitary transformation on the two spatial modes
\begin{align}
\hat{a}^\dag_{s l} \mapsto \sum_{t} \hat{a}^\dag_{t l} U_{\scriptsize\young(t), \scriptsize\young(s)} . \label{eq:fund}
\end{align}
Here we have denoted the fundamental (single photon) $2 \times 2$ matrix representation of the unitary group U(2) by the Young diagram~\cite{Hamermesh} ${\tiny\yng(1)}$ for consistency with what follows; $\left( U^{\tiny\yng(1)} \right)_{ts} = U_{\scriptsize\young(t), \scriptsize\young(s)}$ are the elements of what is sometimes called the transfer matrix in optics.
The action of an interferometer on a complete System$\otimes$Label state is given by $U^{\tiny\yng(1)} \otimes \Id$.
We would like to decompose Fock states in this product basis; this is known as a Schmidt decomposition in quantum information~\cite{NC}, while for doubly indexed creation operators in many-body physics it is the purview of unitary-unitary duality~\cite{RCR2012}.

It will be useful to see how second quantized states are related to first quantized single particle states (e.g.~\cite{Sandu}).
Viewing each excitation of our four mode aggregate as a particle with four available states ($1R, 1G, 2R, 2G$), and recognizing that as bosons the total state must be symmetric under the action of the permutation group S$_2$ that exchanges particles, we have a one-to-one relationship between the Fock states of two bosons in four modes and totally symmetric states of two four-dimensional particles, (often called qu$d$its, here with $d=4$).
Writing the totally symmetric first quantized states in the System$\otimes$Label basis, Eqs.~(\ref{eq:i},\ref{eq:d}) become
\begin{align}
\ket{\psi_\mr{i}} &= \Ket{\begin{matrix} 1&0\\1&0 \end{matrix}} =\frac{1}{\sqrt{2}}\left( \ket{12}+\ket{21} \right) \otimes \ket{RR} , \label{eq:1stind}\\
\ket{\psi_\mr{d}} &= \Ket{\begin{matrix} 1&0\\0&1 \end{matrix}} = \frac{1}{\sqrt{2}} \ket{12} \otimes \ket{RG}+ \frac{1}{\sqrt{2}} \ket{21} \otimes \ket{GR} . \label{eq:1stdis}
\end{align}
Clearly Eq.(\ref{eq:1stind}) is in a product state (Schmidt rank 1) of System and Label\footnote{The System state itself is entangled however, in the (impractical) first quantized sense -- this has been referred to as ``free'' entanglement by Aaronson.}, so the Label states are uncorrelated to the System states; learning the Label does not allow one to learn anything about the System, as one would expect for indistinguishable particles.
Equation (\ref{eq:1stdis}) is entangled (Schmidt rank 2), with the System states perfectly correlated to the Labels ($1\leftrightarrow R$ and $2\leftrightarrow G$), making the photons completely distinguishable.
This entanglement gives rise to mixed states and loss of coherence when the Label is ignored (traced out)~\cite{Schlosshauer}.

Unitary-unitary duality gives a general formalism for this situation.
Passive (particle conserving) operations on this System are given by elements of U(2) since there are two spatial modes $(s=1, 2)$; similarly, one could consider passive operations on the Label, which in this case would also be given by U(2) transformations for the two orthogonal Label states $(l=R,G)$.
Together these actions define transformations in the product group $\mr{U}(2) \times \mr{U}(2)$ that are `local' in that they affect each degree of freedom separately; in particular, they cannot create correlations between the System and Label.
The full set of (passive) `global' transformations on both degrees of freedom is in this case given by $\mr{U}(4) \supset \mr{U}(2) \times \mr{U}(2)$.
Mathematically, the duality theorem states that a representation of this U(4) can be decomposed into the irreducible representations (irreps) of the System's U(2), and that these are in one-to-one correspondence with the irreps of the Label's U(2).
Practically this simply means that there is a change of basis that block diagonalizes the multi-photon representation of the interferometer's action $U^{\tiny\yng(1)}\otimes\Id$, and that each block is indexed by an irrep which we'll denote by $\lambda$.
Note that even though we are discussing a system of bosons and therefore deal only with the totally symmetric irrep of U(4), irreps other than the totally symmetric one occur for the subgroup U(2) $\times$ U(2), and any spatial interferometer's action is by definition an element of this subgroup.

The irreps that occur are given by Young diagrams with as many boxes as particles; in this case $\lambda=\tiny\yng(2), \tiny\yng(1,1)$.
For U(2) these two irreps are well known; they are (for arbitrary single particle quantum numbers $x,y$) the symmetric triplet 
\begin{align}
\Ket{\scriptsize\young(xx)\,} &= \ket{xx} , \\
\sqrt{2}\Ket{\scriptsize\young(xy)\,} &= \ket{xy} + \ket{yx} , \\
\Ket{\scriptsize\young(yy)\,} &= \ket{yy} ,
\end{align}
and the antisymmetric singlet 
\begin{align} 
\sqrt{2}\Ket{\scriptsize\young(x,y)} &= \ket{xy} - \ket{yx} .
\end{align}
We can now rewrite Eqs.~(\ref{eq:1stind}, \ref{eq:1stdis}) as
\begin{align}
\ket{\psi_\mr{i}} = \Ket{\begin{matrix} 1&0\\1&0 \end{matrix}} &= \Ket{\scriptsize\young(12)\,} \Ket{\scriptsize\young(RR)\,} \label{eq:ind} , \\ 
\ket{\psi_\mr{d}} = \Ket{\begin{matrix} 1&0\\0&1 \end{matrix}} &= \frac{1}{\sqrt{2}} \Ket{\scriptsize\young(12)\,} \Ket{\scriptsize\young(RG)\,} 
+ \frac{1}{\sqrt{2}} \Ket{\scriptsize\young(1,2)} \Ket{\scriptsize\young(R,G)} . \label{eq:dis}
\end{align}
Note that total exchange symmetry is indeed preserved because the System and Label states in the second term of Eq.(\ref{eq:dis}) are both antisymmetric.
We see that in general distinguishability is associated to states with amplitude in more than one irrep of the local unitary-unitary group~\cite{ATMS2008,QCMC2010}, as this gives higher Schmidt rank and correlations (i.e. entanglement) in the first quantized representation. 

This unitary-unitary basis provides us with a Schmidt decomposition of the second quantized Fock arrays, while simultaneously block diagonalizing the action of the interferometer.
This means that the System states transform according to matrices $U^{\tiny\yng(2)}$ and $U^{\tiny\yng(1,1)}$, the two-photon irreps corresponding to the fundamental one-photon matrix representation $U^{\tiny\yng(1)}$ of Eq.(\ref{eq:fund}).
Looking at Eqs.~(\ref{eq:ind}, \ref{eq:dis}), we are interested in the matrix elements 
\bea
U_{\tiny\young(11), \tiny\young(12)} &=& \sqrt{2} \, U_{\tiny\young(1), \tiny\young(1)} \, U_{\tiny\young(2), \tiny\young(1)} \label{eq:U2} , \\
U_{\tiny\young(12), \tiny\young(12)} &=& U_{\tiny\young(1), \tiny\young(1)} \, U_{\tiny\young(2), \tiny\young(2)} + U_{\tiny\young(1), \tiny\young(2)} \, U_{\tiny\young(2), \tiny\young(1)} \label{eq:per} , \\
U_{\tiny\young(22), \tiny\young(12)} &=& \sqrt{2} \, U_{\tiny\young(1), \tiny\young(2)} \, U_{\tiny\young(2), \tiny\young(2)} , \label {eq:per2}\\
U_{\tiny\young(1,2), \tiny\young(1,2)}  &=& U_{\tiny\young(1), \tiny\young(1)} \, U_{\tiny\young(2), \tiny\young(2)} - U_{\tiny\young(1), \tiny\young(2)} \, U_{\tiny\young(2), \tiny\young(1)}  , \label{eq:det}
\eea
where we've used a Clebsch-Gordan transformation to perform the block diagonalisation of $U^{\tiny\yng(1)} \otimes U^{\tiny\yng(1)}$ into $U^{\tiny\yng(2)} \oplus U^{\tiny\yng(1,1)}$; analogous transformations exist for any number of particles in any number of modes (e.g.~\cite{GenClebsch}).
We see that the two-photon representations are built from permanents (Eqs.~(\ref{eq:U2}) - (\ref{eq:per2})) and determinants (Eq.~(\ref{eq:det})) of the representation $U^{\tiny\yng(1)} \otimes \Id$~\cite{Scheel2004, AA, TGdGS}.

A balanced beamsplitter is given by the unitary $B^{\tiny\yng(1)} = \frac{1}{\sqrt{2}} \left[ \begin{matrix} 1&i\\i&1 \end{matrix} \right] $ and according to the above acts on the System's Schmidt basis as
\begin{align}
2 B^{\tiny\yng(2)} \Ket{\scriptsize\young(11)\,} &= \Ket{\scriptsize\young(11)\,}+{i}{\sqrt{2}}\Ket{\scriptsize\young(12)\,}-\Ket{\scriptsize\young(22)\,} , \label{eq:B1} \\
\sqrt{2} B^{\tiny\yng(2)} \Ket{\scriptsize\young(12)\,} &= i \Ket{\scriptsize\young(11)\,}+ i \Ket{\scriptsize\young(22)\,} , \label{eq:hom} \\ 
2 B^{\tiny\yng(2)} \Ket{\scriptsize\young(22)\,} &= - \Ket{\scriptsize\young(11)\,}+ {i}{\sqrt{2}}\Ket{\scriptsize\young(12)\,}+\Ket{\scriptsize\young(22)\,} , \label{eq:B3}\\
B^{\tiny\yng(1,1)} \Ket{\scriptsize\young(1,2)} &=  \Ket{\scriptsize\young(1,2)} . \label{eq:inv}
\end{align}
Assuming our detectors are insensitive to the Label (here frequency), a coincidence count in System modes 1 and 2 is given by the sum of projectors onto the coincident spatial states for both the triplet and singlet:
\be
C_{12} = \left( \Ket{\scriptsize\young(12)\,}\Bra{\scriptsize\young(12)} + \Ket{\scriptsize\young(1,2)\,}\Bra{\scriptsize\young(1,2)\,} \right) \otimes \Id . \label{eq:coin}
\ee

We can now see how the HOM effect occurs in this picture.
Equation (\ref{eq:per}) indicates that the triplet matrix element taking the coincident (one photon in each spatial mode) input to the coincident output is the permanent of $U^{\tiny\yng(1)}$, and Eq.(\ref{eq:hom}) shows that for the balanced beamsplitter this permanent disappears.
As the indistinguishable state in Eq.(\ref{eq:ind}) has only a triplet component, it gives no coincidences.
The distinguishable state in Eq.(\ref{eq:dis}) however has amplitude in the singlet subspace of spatial modes 1 and 2, which Eq.(\ref{eq:det}) shows transforms as the determinant.
The beamsplitter has determinant 1 so this amplitude is unchanged, as shown by Eq.(\ref{eq:inv}), and the second term in Eq.(\ref{eq:coin}) indicates that the surviving singlet amplitude gives rise to coincidence counts.
A HOM dip occurs by preparing a superposition of these two states and varying the amplitudes from Eq.(\ref{eq:dis}) to Eq.(\ref{eq:ind}) and back again.
Note that this requires direct manipulation of the Label degree of freedom, something we'd like to avoid as it is usually not possible in practice; in order for a two photon state with amplitude in the singlet subspace to exhibit no coincidences, the determinant of the transfer matrix must be zero -- obviously this would violate unitarity for two System modes.

\begin{figure}
\begin{tikzpicture}[scale=1]
 \draw [thick] (0,2) -- (1,2);
 \draw [thick] (0,1) -- (1,1);
 \draw [thick] (0,0) node [left] {$\ket{\mathrm{vac}}$} -- (2,0);
 \draw (-1.1,1.5) node {$\Ket{\scriptsize\young(RG)\,}, \Ket{\scriptsize\young(R,G)\,} \Bigg\{$};
 \draw [thick] (1,2) -- (3,0);
 \draw [thick] (1,1) -- (2,2);
 \draw [thick] (2,0) -- (3,1);
 \draw [thick,dashed] (1.2,1.5) -- (1.8,1.5);
 \draw [thick,dashed] (2.2,0.5) -- (2.8,0.5);
 \draw [thick] (2,2) -- (4,2) node [right] {$\bra{\mathrm{vac}}$};
 \draw [thick] (3,1) -- (4,1);
 \draw [thick] (3,0) -- (4,0);
 \draw (4.7,0.5) node {$\Bigg\} \Ket{\scriptsize\young(RG)\,}$};
\end{tikzpicture}
\caption{
An interferometer corresponding to Eq.(\ref{eq:ElBuncherino}), consisting of two balanced beamsplitters.
When two coincident photons bunch at the first beamsplitter and yield vacuum at the top output, their spatial System state no longer has an antisymmetric component.
This makes it impossible for the photons' Label modes $R,G$ to be correlated with their System modes, rendering them indistinguishable.
A subsequent beamsplitter is a local transformation on the System and cannot create correlations, allowing the (now purely symmetric) output in the bottom two ports to pass a HOM test.
\label{fig:ElB}}
\end{figure}
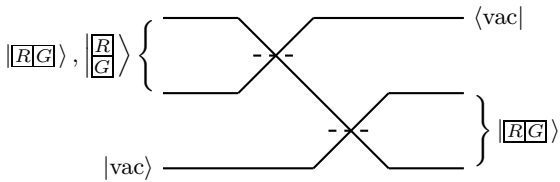

We now show that such unitaries do exist for three modes.
Consider a partially distinguishable pure state of two photons in System modes 1 \& 2
\begin{align}
\ket{\psi} &= \alpha\Ket{\begin{matrix} 1&0\\1&0\\0&0 \end{matrix}} + \beta\Ket{\begin{matrix} 1&0\\0&1\\0&0 \end{matrix}} \label{eq:imphom} \\
&= \Ket{\scriptsize\young(12)\,} \left( \alpha \Ket{\scriptsize\young(RR)\,} + \frac{\beta}{\sqrt{2}} \Ket{\scriptsize\young(RG)\,} \right)
 + \Ket{\scriptsize\young(1,2)\,} \left( \frac{\beta}{\sqrt{2}} \, \Ket{\scriptsize\young(R,G)\,} \right) ,
\end{align}
where $|\alpha|^2+|\beta|^2=1$.
The partial distinguishability ($\alpha, \beta \neq 0$) manifests as Label states that are neither perfectly correlated nor uncorrelated to the System states\footnote{Including the other two possible Fock arrays in the superposition does not materially change the analysis.}.
An interferometer's transfer matrix $U^{\tiny\yng(1)}$ is now an element of U(3), and let us assume that our HOM test, a balanced beamsplitter, will be made on System output modes 2 \& 3, (any pair of outputs can be chosen, but as we'll see this yields a particularly simple interferometer).
The submatrix taking inputs 1 \& 2 to outputs 2 \& 3 is given by $\left( U^{\tiny\yng(1)} \right)_{2 \cdots 3,1 \cdots 2}$ -- as argued above, we need its determinant to vanish.
A solution is given by
\begin{align}
U^{\tiny\yng(1)} = \frac{1}{2} \left[ \begin{matrix} \sqrt{2} & \sqrt{2}i & 0 \\ i & 1 & \sqrt{2}i \\ -1 & i & \sqrt{2} \end{matrix} \right] . \label{eq:ElBuncherino}
\end{align}
Postselecting vacuum in mode 1, the method employed above for two modes can be used to compute the unnormalized output state
\begin{align}
\frac{1}{2\sqrt{2}} \left( i \Ket{\scriptsize\young(22)\,} - \sqrt{2} \Ket{\scriptsize\young(23)\,} - i \Ket{\scriptsize\young(33)\,} \right)
\left( \alpha \Ket{\scriptsize\young(RR)\,} + \frac{\beta}{\sqrt{2}} \Ket{\scriptsize\young(RG)\,} \right) , \label{eq:out}
\end{align}
whose norm gives the postselection probability $\frac{1}{2}\left(1-|\beta|^2/2\right)$.
Replacing mode 1 with 2 and 2 with 3 in Eqs.~(\ref{eq:B1}-\ref{eq:B3}, \ref{eq:coin}) we see that a subsequent beamsplitter in modes 2 \& 3 does indeed yield no coincidences.
Roughly speaking, this unitary allows the two photons to emerge in output modes 2 \& 3, but `filters' the (anti)symmetry~\cite{ASS2007} of the input such that when they do their spatial state is purely symmetric, of Schmidt rank 1, and therefore uncorrelated with the Label states.
At the same time, the noncoincident spatial terms ({\scriptsize\young(22)} and {\scriptsize\young(33)}) at the output will interfere destructively on a balanced beam splitter.
This ensures that the output passes a HOM test, insofar as it gives no coincidences.
As is often the case in linear optics, the price paid is nondeterminism; distinguishable states successfully pass with nonzero probability, but a completely indistinguishable state ($\beta=0$) now only does so half the time.
Figure~\ref{fig:ElB} shows a realization of this interferometer.

Should this be interpreted as quantum interference of distinguishable particles?
From Eqs.(\ref{eq:fund}, \ref{eq:ElBuncherino}), we have
\begin{align}
2 \hat{a}^\dag_{1l} &\stackrel{U}{\mapsto} \sqrt{2}\hat{a}^\dag_{1l} + i \hat{a}^\dag_{2l} - \hat{a}^\dag_{3l} ,\\
2 \hat{a}^\dag_{2l} &\stackrel{U}{\mapsto} \sqrt{2}i\hat{a}^\dag_{1l} + \hat{a}^\dag_{2l} + i\hat{a}^\dag_{3l} .
\end{align}
Classically this implies that a photon input into System mode 1 emerges from modes 1, 2, 3 with probabilities 1/2, 1/4, 1/4 respectively.
The same occurs for a photon input into mode 2, and in both cases the photons' Labels $l$ are unaffected.
Thus we expect that photons should emerge from modes 2 \& 3 coincidently $1/4\cdot1/4 + 1/4\cdot1/4 = 1/8$ of the time\footnote{Classical postselection on no photons emerging from output 1 serves only to renormalize these probabilities.}.
The quantum analysis agrees, as expected for distinguishable particles.
As discussed in the introduction, the classical interpretation of a subsequent balanced beamsplitter concludes that there should be coincidences at the output of the beamsplitter 1/16 of the time.
However, as explained above, Eq.~(\ref{eq:out}) implies that the quantum calculation yields exactly zero coincidences, and so in the same way as the HOM effect we are led to conclude that, despite the fact that our input was distinguishable, this is indeed quantum interference.
To muddy the waters, compare this with the overall unitary that includes a final HOM beamsplitter in modes 2 \& 3:
\begin{align}
B^{\tiny\yng(1)} U^{\tiny\yng(1)}=\frac{1}{\sqrt{2}} \left[ \begin{matrix} 1 & i & 0 \\ 0 & 0 & \sqrt{2}i \\ -1 & i & 0 \end{matrix} \right] . 
\end{align}
The classical interpretation of this transformation is also in agreement with the quantum; not only are there no coincidences in outputs 2 \& 3, there are no photons in output 2 at all.
This can be seen from Fig.~\ref{fig:ElB}; a third balanced beamsplitter in modes 2 \& 3 serves to simply swap the vacuum input from mode 3 into mode 2.
What this shows is that the distinguishability was already erased after the first beamsplitter and postselection; adding the third mode simply allows us to perform a HOM test.

We reiterate that unlike the original HOM experiment, here we act only trivially on the Label -- the action is manifestly local, so indeed we need not assume anything about this degree of freedom.
We might therefore interpret Eq.(\ref{eq:ElBuncherino}) as a kind of filter~\cite{Birchall}, in that if we postselect on vacuum in System mode 1 -- a nonlocal operation -- we are left with a product state regardless of how distinguishable the input was, (this can be viewed from a quantum information perspective as an example of decoupling~\cite{Dupuis}).
The output is independent of the amplitude of the distinguishability $\beta$, as changing its value affects only the postselection probability.
The effect is robust in that there is a continuum of such unitaries, with
\begin{align}
&U^{\tiny\yng(1)} (\theta,\phi,\xi,\zeta) = \nonumber \\
&\frac{1}{\sqrt{2}} \left[ \begin{matrix} 
-e^{i\zeta} & -ie^{i\zeta} & 0 \\
ie^{-i(\phi+\zeta)}\cos\theta & e^{-i(\phi+\zeta)}\cos\theta & -\sqrt{2}e^{-i\xi}\sin\theta  \\ 
ie^{i(\xi-\zeta)}\sin\theta & e^{i(\xi-\zeta)}\sin\theta & \sqrt{2}e^{i\phi}\cos\theta
\end{matrix} \right] 
\end{align}
giving the same result for $\theta \neq 0,\pi/2$.
For applications we would like the noncoincident terms in Eq.~(\ref{eq:out}) to also be filtered, however one can show that it is impossible to make precisely one of either the permanent or determinant of a submatrix as well as the noncoincident terms vanish simultaneously, even for arbitrarily many modes.

In conclusion, we've seen that unitary-unitary duality in a first quantized approach gives a natural Schmidt decomposition for multimode states, and that entangled correlations correspond to distinguishability.
Interferometers that act on only one of the Schmidt factors -- the ``System'' -- cannot break this entanglement, but they can shift it around in such a way that subsequent postselection does, yielding a product state.
This can (probabilistically) render a completely distinguishable state completely indistinguishable, without any manipulation of the distinguishing ``Label'' degree(s) of freedom.

The real strength of this formalism is that it accommodates any number of System modes $S$, Label modes $L$, and particles $N$, where we consider the subgroup chain U($SL$) $\supset$ U($S$) $\times$ U($L$).
This makes it useful for dealing with distinguishability in increasingly complex interference experiments, (e.g.~\cite{Tillmann}), and large scale photonic quantum information processing, such as Boson Sampling~\cite{BosSam}.
In general many irreps $\lambda$ occur, and immanants can be used to construct the $N$-particle representations of the unitary matrices similarly to the use of permanents and determinants here~\cite{TGdGS,dG}.
Complications that arise are the occurrence of multiplicities of irreps for $N>2$, necessitating a sum over the multiplicity index in order to construct totally symmetric states in the Schmidt decomposition, as well as multiplicities of irrep weights which can occur for U($d>2$).

The approach is not only applicable to bosons; unitary-unitary duality can be viewed as a coupling problem between states of the Schur-Weyl decompositions of the two factors in the Schmidt decomposition \cite{ATMS2008, RCR2012}.
Here we are coupling to the totally symmetric representation of U($SL$), (e.g. Eq.~(\ref{eq:dis})), however coupling to other irreps is also possible; in particular coupling to the totally antisymmetric gives the analogous picture for fermions.

\emph{Acknowledgements --}
The author would like to thank R.~Adamson, S.~Bartlett, P.~Birchall, H.~de~Guise, A.~Laing, D.~Mahler, J.~Matthews, M.~Murao, D.~Rowe, T.~Rudolph, N.~Russel, B.~Sanders, C.~Sparrow, S.~Stanisic, A.~Steinberg and S.~Takanori for helpful discussions on this topic over the years.
This effort is supported in part by the U.S. Army Research Office under contract W911NF-14-1-0133 and EPSRC First Grant EP/N014812/1.

\end{document}